\documentclass[epsfig ,12pt]{article}
\usepackage[dvips]{graphicx}
\usepackage{amsmath,amssymb,latexsym}
\newcommand{\ro}{\langle\rho\rangle}
\newcommand{\rom}{\langle\rho_{min}\rangle }
\newcommand{\robm}{\langle\bar{\rho}_{min}\rangle }
\begin{document}
\begin{titlepage}
\begin{center}
\vfill
{\large\bf Quantum interest for scalar fields in Minkowski spacetime}
\vfill
\end{center}

\begin{center}
Frans Pretorius 
\end{center}
\vfill

\begin{center}{\sl
Department of Physics and Astronomy \\
University of Victoria \\
P.O. Box 3055 STN CSC \\
Victoria B.C., Canada V8W 3P6 \\
}\end{center}
\vfill

\begin{abstract} 
The quantum interest conjecture of Ford and Roman states that any negative 
energy flux in a free quantum field must be preceded or followed by a 
positive flux of greater magnitude, and the surplus of positive energy 
grows the further the positive and negative fluxes are apart. In addition, 
the maximum possible separation between the positive and negative energy 
decreases the larger the amount of negative energy. We prove that the 
quantum interest conjecture holds for arbitrary fluxes of non-interacting 
scalar field energy in 4D Minkowski spacetime, and discuss the 
consequences in attempting to violate the second law of thermodynamics 
using negative energy. We speculate that quantum interest may also hold 
for the Electromagnetic and Dirac fields, and might be applied to certain 
curved spacetimes.
\end{abstract}

\vfill

\end{titlepage}

\setcounter{page}{2}
\section{Introduction}

Quantum field theory permits the existence of states where the 
renormalized energy density can become arbitrarily negative in regions of 
spacetime even though the total energy is always positive 
\cite{neg_in_qft}. Negative energy is an essential ingredient in many 
bizarre effects, including wormholes \cite{worm}, warp drives \cite{warp}, 
time machines \cite{time}; and may be used to violate the $2^{nd}$ law of 
thermodynamics \cite{2ndlaw_ford}, \cite{2ndlaw_davies}. Fortunately (or 
unfortunately!) there appear to be severe restrictions on the magnitude 
and duration of negative energies that might occur in a quantum field. One 
form of these restrictions are the ``quantum inequalities'', originally 
proposed by Ford and Roman \cite{ford_qi}, \cite{ford_roman_qi} and 
studied by numerous authors since \cite{other_qi}, which essentially state 
that large amounts of negative energy can only be ``seen'' for very short 
intervals of time. These inequalities have been used to place stringent 
limitations on warp drive and wormhole geometries 
\cite{warp_res},\cite{worm_res}.  \par
\begin{figure}
\begin{center}
\includegraphics[viewport =28 460 372 730, clip] {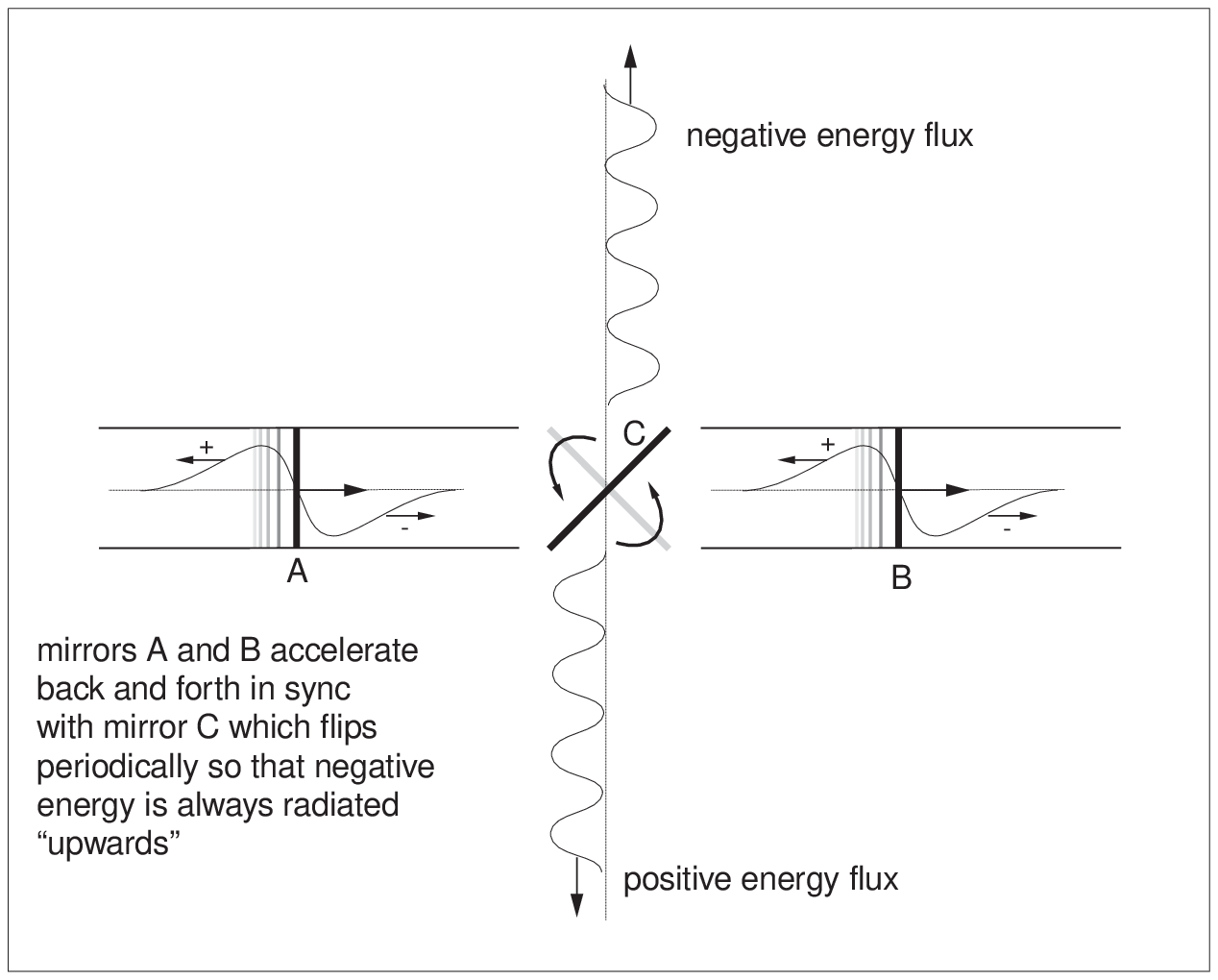}
\end{center}
\caption{\label{fig_mirror} A pair of mirrors (A and B) accelerate back 
and forth in tandem, radiating negative energy in the direction of 
increasing acceleration and positive energy in the opposite direction. 
Between these two mirrors is a third mirror (C) that periodically flips so 
that the negative energy parts of the flux generated by mirrors A and B 
are always reflected in the same direction. Such a device could in 
principle produce a steady flow of negative energy, \emph{if} it were 
possible to reflect just the negative part of the mirror flux.}
\end{figure}

Recently, Ford and Roman proposed the ``quantum interest conjecture'' and 
proved it for delta function pulses of negative energy for massless scalar 
fields in 2D and 4D Minkowski spacetime \cite{q_int}. This conjecture is a 
consequence of the quantum inequalities (QI's), and states that any 
negative energy pulse (the ``loan'') must be accompanied (``repaid'') by a 
positive energy pulse within a certain maximum time interval, and the 
larger the separation of the pulses the larger the magnitude the positive 
pulse must be relative to the negative pulse (i.e., repaid with 
``interest''). At first glance this statement may not seem too profound -- 
after all the total energy must be positive, so if there is a location 
with negative energy there will be compensating positive energy somewhere 
in the spacetime. But the quantum interest conjecture tells us a lot more 
about the nature of negative energies in free-fields: negative energy is 
always in close proximity to an entourage of positive energy. This, for 
instance, has immediate consequences in attempts to violate the $2^{nd}$ 
law of thermodynamics. For suppose negative energies were ``substantial'' 
enough that one could in principle reflect only the negative energy part 
of the flux produced by an accelerating mirror as shown in Figure 
\ref{fig_mirror} (a variant of a device first proposed by Davies 
\cite{2ndlaw_davies} who used it to construct a reversible process that 
effectively transferred energy from a cold body to a hot one without doing 
work). The resultant stream of negative energy could be sent far enough 
away from the device so that one could reasonably apply the free-field 
quantum inequalities to the stream. Even though each pulse within the 
stream may be consistent with the original quantum inequality, the 
stronger quantum interest conjecture strictly forbids such a flux of 
negative energy. This implies that the mirror device in Figure 
\ref{fig_mirror} cannot exist; if we want to reflect negative energy we 
must reflect its support of positive energy, which is at least as large in 
magnitude. Thus one cannot subject a hot body to a pure flux of negative 
energy to lower its entropy (at least using scalar quantum fields), as 
suggested in \cite{2ndlaw_davies}.\par
In this paper, using a simple scaling argument, we present a proof of 
quantum interest for arbitrary distributions of negative energy of scalar 
fields in 4D Minkowski spacetime (slightly weaker results are obtained in 
2D Minkowski spacetime). We do this first for the massless scalar field in 
section \ref{sec_massless}, after introducing the quantum inequalities in 
section \ref{sec_qi}. In section \ref{sec_massive} we show that a massive 
scalar field has stronger constraints on the magnitude and duration of 
negative energies than a massless field, thus making the results of 
section \ref{sec_massless} applicable to both types of scalar field. In 
section \ref{sec_other} we briefly comment on the possibility of extending 
quantum interest to the Electromagnetic and Dirac fields, curved 
spacetimes and to situations in Minkowski space where mirror-like boundary 
conditions are imposed on the fields.

\section{\label{sec_qi} Quantum inequalities}

The quantum inequalities can be stated as follows. An inertial observer 
samples the local energy density $\rho(t)$ over a period of time with a 
sampling function $g(t)$ to obtain an average energy density $\ro$:
\begin{equation}
\ro=\int_{-\infty}^{\infty} g(t) \rho(t) dt.
\end{equation}
The only conditions imposed upon $g(t)$ are that
\begin{equation}
\int_{-\infty}^{\infty} g(t)dt=1, \ \ \ \mbox{and} \ \ \ g(t)\ge 0 \ \ 
\forall t.
\end{equation}
Then,
\begin{equation}\label{qi}
\ro \ge \rom,
\end{equation}
where $\rom$ is a constant that depends upon the sampling function $g(t)$ 
and the dimensionality $d$ of the spacetime. Note that for a given energy 
density $\rho(t)$ (\ref{qi}) must be satisfied by \emph{all} choices of 
$g(t)$. Flanagan's optimal bound for a massless scalar field in $2D$ is 
\cite{flanagan}
\begin{equation} \label{2df}
\rom=-\frac{1}{24\pi} \int_{-\infty}^{\infty} \frac{g'(t)^2}{g(t)}dt,
\end{equation}
while Fewster and Eveson obtained the following bounds in 2D and 4D 
Minkowski spacetime \cite{fe_flat}:
\begin{eqnarray}  
\rom=-\frac{1}{16\pi} \int_{-\infty}^{\infty} \frac{g'(t)^2}{g(t)}dt, \ \ 
\  \label{2def}  \mbox{(2D)} \\
\rom=-\frac{1}{16\pi^2} \int_{-\infty}^{\infty} 
\left(g^{1/2}(t)''\right)^2dt, \ \ \  \label{4def} \mbox{(4D)}.
\end{eqnarray}
Certain sampling functions will not give a lower bound, in particular if 
there are discontinuities in $g(t)$ or $g'(t)$. For example the 
rectangular pulse function ($g(t)=\frac{1}{t_0}$ when $-\frac{t_0}{2} < t 
< \frac{t_0}{2}$ and $0$ elsewhere) doesn't give a finite lower bound 
$\rom$. This makes sense if we recall the positive/negative energy delta 
pulse pair produced by a mirror that instantaneously accelerates from rest 
(producing a negative pulse), then, after undergoing a period of uniform 
acceleration, decelerates to zero acceleration (emitting a positive pulse) 
\cite{q_int}. The magnitude of energy produced by the mirror is 
proportional to its change in acceleration with time. We can thus make the 
negative pulse as energetic as we want, but doing so shortens the time 
interval before the positive pulse arrives (the mirror is decelerated 
before it crashes into the observer). If we sample the negative energy 
with the rectangular function we can avoid measuring any positive energy 
by timing the rectangular function to turn off before the positive pulse 
arrives. \par
More insight into the intimate relationship between the sampling function 
and minimum bound can be obtained from the derivation of Fewster and 
Eveson. One can write (\ref{4def}) as \cite{fe_flat}
\begin{equation}
\rom=-\frac{1}{16\pi^3}\int_0^\infty (\widehat{g^{1/2}}(w))^2 w^4 dw,
\end{equation}
where $\widehat{g^{1/2}}(w)$ is the Fourier transform of the square root 
of $g(t)$. Smooth sampling functions, like the Lorentzian function 
originally employed by Ford, decay rapidly in the frequency domain, 
smoothing over higher frequency (hence higher energy) transient components 
of the flux. Negative energies in a free field appear to be coherence or 
interference effects produced by peculiar superpositions of the positive 
mode quanta of the field. For example, the well-known vacuum + 2 particle 
state $|\psi \rangle = \alpha |0\rangle +\beta|2\rangle $ has negative 
energy at periodic intervals with appropriate choices of $\alpha$ and 
$\beta$: the frequency and energy density of the negative regions are 
proportional to the frequency of the 2-particle modes \cite{other_qi}. 
This suggests that if we want to see a lot of negative energy we need to 
look at such high frequency transient phenomena, and the only way to 
``catch'' the negative energy is to use a sampling function with steep 
edges. But as discussed in the introduction the quantum interest 
conjecture seems to say that one cannot interact with this negative energy 
as one can with positive energy -- ``catch'' may be an overstatement. \par

\section{\label{sec_massless} Quantum interest for massless scalar fields}

The key to obtaining useful information from the quantum inequalities in 
light of the arbitrariness of the sampling function, and hence lower 
bound, is to choose an appropriate class of sampling function. To prove 
quantum interest, we will use a function $g(t)$ with compact support 
($g(t)$ is zero outside the range $[-t_0/2,t_0/2]$), that has a single 
maximum at $t=0$ and is sufficiently smooth such that a lower bound in 
(\ref{2df}) - (\ref{4def}) exists. For simplicity we will also assume that 
$g(t)$ is symmetric about $t=0$. For example, the following sampling 
functions will do (though for the most part the particular choice won't 
matter):
\begin{eqnarray}\label{cos}
g(t) \propto \cos^n\left(\frac{\pi t}{t_0}\right), \ \ \ -
\frac{t_0}{2}<t<\frac{t_0}{2}, \ \ \ (n\ge 2) \\
0 \ \ \ elsewhere, \nonumber
\end{eqnarray}
or
\begin{eqnarray}\label{power}
g(t) \propto \left(t^2-\frac{t_0^2}{4}\right)^n, \ \ \ -
\frac{t_0}{2}<t<\frac{t_0}{2}, \ \ \ (n\ge 2) \\
0 \ \ \ elsewhere. \nonumber
\end{eqnarray}
The minimum bounds are strongest (least negative) when $n=2$; as $n 
\rightarrow \infty$ these functions approach $\delta(t)$ which has no 
lower bound.\par
\begin{figure}
\begin{center}
\includegraphics[viewport =28 510 510 740, clip] {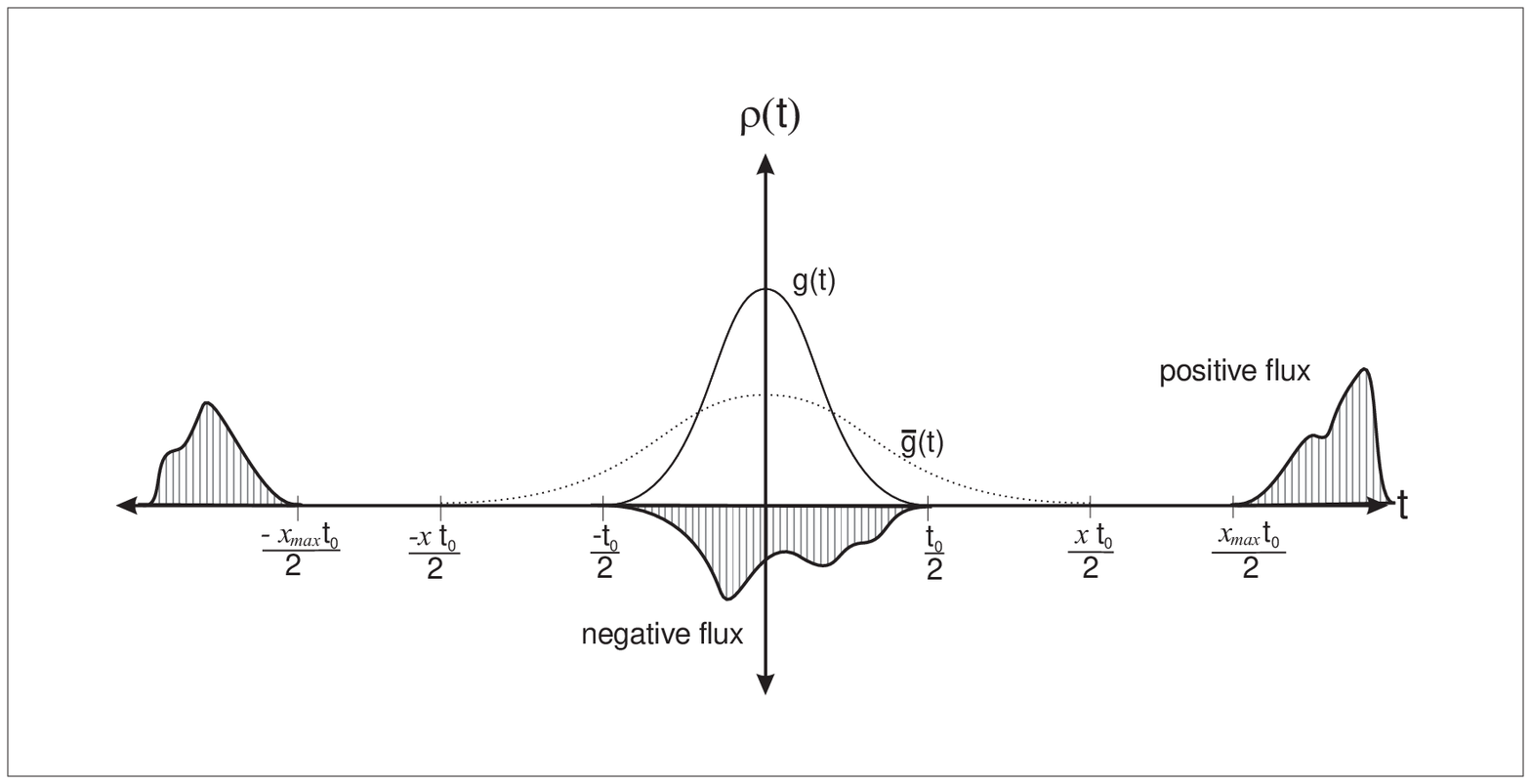}
\end{center}
\caption{\label{fig_flux1} We sample a local distribution of negative 
energy with a function $g(t)$ and with a scaled version of $g(t)$, namely 
$\bar{g}(t)=g(t/x)/x$. The quantum inequalities then tell us that there is 
a maximum scale factor $x_{max}$ beyond which positive energy \emph{must} 
be sampled by $\bar{g}(t)$.}
\end{figure}

Now consider the hypothetical situation shown in Figure (\ref{fig_flux1}). 
We have an isolated distribution of negative energy flowing past the 
observer who samples it with a function $g(t)$ like (\ref{cos}) or 
(\ref{power}), timed to snugly encompass the negative flux. We want to 
answer two questions:

1) How isolated can the negative pulse be? In other words, how soon before 
or after the negative flux arrives \emph{must} one see positive energy. 

2) When we do start sampling positive energy, must one pay quantum 
interest? I.e., does the total positive energy outweigh the negative 
energy by an amount that increases the further the two pulses are apart. 
\par
To answer these questions we sample the distribution again with a second 
function $\bar{g}(t)$ that is merely a copy of $g(t)$ scaled by a factor 
$x \ge 1$:
\begin{equation}\label{gb}
\bar{g}(t)=\frac{1}{x}g\left(\frac{t}{x}\right).
\end{equation} 
The support of $\bar{g}(t)$ is thus $[-x t_0/2, x t_0/2]$, and the leading 
factor of $1/x$ is a normalization constant to give $\bar{g}$ unit 
integral. If we calculate the minimum negative energy density $\robm$ 
allowed by the quantum inequalities using $\bar{g}$ in (\ref{2df}) or 
(\ref{2def}) for 2D and (\ref{4def}) in 4D Minkowski spacetime we obtain 
the key result:
\begin{equation}\label{scale}
\robm=\frac{\rom}{x^d}.
\end{equation}
Here $\rom$ is the lower bound associated with $g(t)$ and $d$ is the 
spacetime dimension (2 or 4). This expression immediately suggests the 
principle of quantum interest. We have total negative energy $E_m=\int_{-
t_0/2}^{t_0/2} \rho(t) dt$ and an average energy density of 
$\rho_{avg}=E_m/t_0 \approx \ro$. If we now increase our sampling range to 
$x t_0$, and $\rho(t)$ is zero outside of $[-t_0/2, t_0/2]$, then 
$\rho_{avg}$ will scale as $1/x$. But the maximum allowed negative energy 
density scales as $1/x^{d}$, thus positive energy (and probably quite a 
lot of it) is eventually needed to satisfy the quantum inequalities. \par
We can make the preceding statement more precise. Define a constant $y$, 
with $0 < y \le 1$, such that
\begin{equation}\label{y_def}
\ro=\int_{-t_0/2}^{t_0/2} g(t) \rho(t) dt=y \rom.
\end{equation}
Note that for most sampling functions $g(t)$ there will probably not be 
any quantum state that achieves the minimum ($y=1$). Now stretch $g(t)$ by 
the factor $x > 1$, and to answer the first question we will show that 
there is a largest possible $x=x_{max}$ allowed by the QI's if we assume 
zero energy density outside of the negative pulse, as illustrated in 
Figure \ref{fig_flux1}:
\begin{equation}
\int_{-x t_0/2}^{x t_0/2} \rho(t) \bar{g}(t) dt = \frac{1}{x} \int_{-
t_0/2}^{t_0/2} \rho(t) g(t/x) dt \ge \robm = \frac{\rom}{x^d}.
\end{equation} 
Using (\ref{y_def}) we can rewrite the inequality as
\begin{equation}\label{x_max_ie}
x^{d-1} \le \frac{1}{y} \frac{\int_{-t_0/2}^{t_0/2} \rho(t) g(t) 
dt}{\int_{-t_0/2}^{t_0/2} \rho(t) g(t/x) dt}.
\end{equation}
This clearly shows that if we have some negative energy ($y \neq 0$) then 
there is an upper bound on $x$, for, recalling that $g(t)$ is positive 
with a single peak at $t=0$ so that $g(0) \ge g(t/x) \ge g(t)$, one can 
see that the ratio of the two integrals in (\ref{x_max_ie}) is $\le 1$ 
(but is at least as large as $ \frac{\int_{-t_0/2}^{t_0/2} \rho(t) g(t) dt 
}{g(0) \int_{-t_0/2}^{t_0/2} \rho(t)dt}$). Thus we can write
\begin{equation}\label{x_max}
x_{max}^{d-1} = \frac{1}{y} \frac{\int_{-t_0/2}^{t_0/2} \rho(t) g(t) 
dt}{\int_{-t_0/2}^{t_0/2} \rho(t) g(t/x_{max}) dt}.
\end{equation}
This upper bound depends on the sampling function and in general will 
over-estimate the maximum allowed separation since a real distribution of 
energy must satisfy (\ref{x_max}) for all choices of $g(t)$. \par
Without a specific sampling function or energy distribution we cannot 
reduce (\ref{x_max}) any further, but we can see that the range of 
possible $x$ is most strongly influenced by $y$. If $y=1$ (we have a state 
that actually achieves the minimum allowed by $g(t)$) then the only way 
(\ref{x_max_ie}) or (\ref{x_max}) can be satisfied is if $x=1$; i.e. 
positive energy must \emph{immediately} follow and or precede the negative 
energy. If $y$ is close to zero then $x$ can be large and we can 
approximate the integral in the denominator of (\ref{x_max}) by evaluating 
$g(t/x)$ at $t=0$: 
\begin{equation}
x_{max}^{d-1} \approx \frac{1}{y} \frac{\ro}{g(0) E_m}, \ \ \ 1/y \gg 1. 
\end{equation}
In most situations $\ro/g(0) E_m$ will be a number of order unity. If we 
have a delta function pulse of negative energy centered at $t=0$ (as 
considered by Ford and Roman) we obtain a similar relation
\begin{equation}\label{x_max_delta}
x_{max}^{d-1} = \frac{1}{y}.
\end{equation}
The above expressions (\ref{x_max}) - (\ref{x_max_delta}) all show that 
stronger distributions of negative energy (larger y) are required to be 
close to positive energy (smaller $x_{max}$). Also note that the bound on 
$x$ is stronger in 4 dimensional spacetime. \par
\begin{figure}
\begin{center}
\includegraphics[viewport =28 510 486 740, clip] {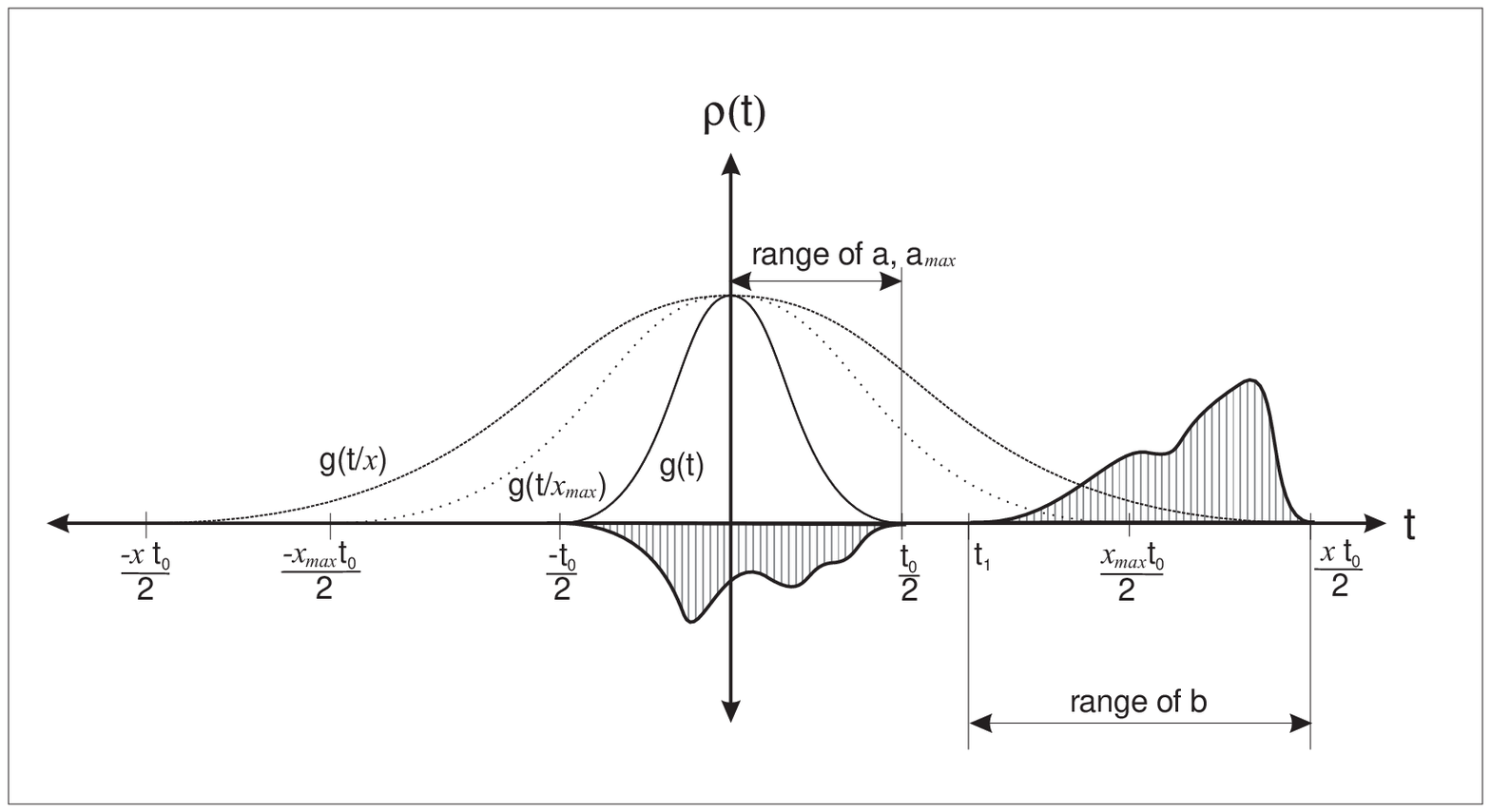}
\end{center}
\caption{\label{fig_flux2} In this situation the positive flux arrives a 
time $\Delta t=t_1-t_0/2$ after the negative pulse, and lasts for a time 
$x t_0/2-t_1$. The quantum inequalities tell us that the total amount of 
positive energy must always be larger than the total amount of negative 
energy.} 
\end{figure}

To answer the second question, namely whether the quantum interest 
$\epsilon$ defined by
\begin{equation}\label{eps_def}
\frac{E_p}{|E_m|}=(1+\epsilon)
\end{equation}
is positive, consider the situation in Figure \ref{fig_flux2} (note that 
in this figure we have omitted the $1/x$ normalization constants in the 
plots of $\bar{g}$), where $E_p$ is the total positive energy, i.e. 
$E_p=\int_{t_1}^{x t_0/2} \rho(t) dt$. Here we stretch $g(t)$ by a new 
factor $x$ (possibly larger than $x_{max}$, which is the maximum $x$ if we 
$\emph{only}$ sample negative energy), and the positive energy flux 
arrives at time $t_1$, with $t_0/2 \le t_1 \le x_{max} t_0/2$. For 
simplicity we only consider positive energy that arrives after the 
negative energy, but this doesn't affect the generality of the argument. 
Applying the QI's and scaling relation to this situation yields
\begin{equation}\label{q_int1}
\int_{-x t_0/2}^{x t_0/2} \rho(t) \bar{g}(t) dt = \frac{1}{x}\int_{-
t_0/2}^{t_0/2} \rho(t) g(t/x) dt+\frac{1}{x}\int_{t_1}^{x t_0/2} \rho(t) 
g(t/x) dt \ge \robm = \frac{\rom}{x^d}
\end{equation}
 
To simplify the appearance of this expression we assume that $\rho(t)$ is 
negative semi-definite in the range $[-t_0/2,t_0/2]$, and positive semi-
definite elsewhere (again this does not qualitatively affect the 
conclusions). Then we can find a number $a$, where $0 \le a < t_0/2$, such 
that  $\int_{-t_0/2}^{t_0/2} \rho(t) g(t/x) dt=g(a/x) E_m$, and a number 
$b$, where $a < t_1 \le b < x t_0/2$, such that $\int_{t_1}^{x t_0/2} 
\rho(t) g(t/x) dt = g(b/x) E_p$ (see Figure \ref{fig_flux2}). Thus we can 
rewrite (\ref{q_int1}) as
\begin{equation}\label{q_int2}
-|E_m| g(a/x) + E_p g(b/x) \ge -\frac{|\rom|}{x^{d-1}}.
\end{equation}
(This expression is already quite suggestive: if the right hand side of 
(\ref{q_int2}) is close to zero then $E_p$ will have to outweigh $|E_m|$ 
by roughly $g(a/x)/g(b/x)$ to satisfy the inequality). Using (\ref{y_def}) 
and (\ref{x_max}), we can write (\ref{q_int2}) as 
\begin{equation}\label{q_int3}
-|E_m| g(a/x) + E_p g(b/x) \ge \left( \frac{x_{max}}{x} \right)^{d-1} 
\int_{-t_0/2}^{t_0/2} g(t/x_{max})\rho(t) dt.
\end{equation}
As we did for (\ref{q_int1}), we can simplify (\ref{q_int3}) using 
$\int_{-t_0/2}^{t_0/2} g(t/x_{max})\rho(t) dt = -g(a_{max}/x_{max}) 
|E_m|$, where $0\le a_{max} < t_0/2$ (note that $a_{max}$ simply labels 
the evaluation of the integral when $x=x_{max}$ and doesn't refer to any 
maximization of the label $a$ defined earlier; in particular, because 
$g(t)$ decreases monotonically away from $t=0$, $x<x_{max}$ implies that 
$a>a_{max}$ and hence $g(a/x)<g(a_{max}/x_{max})$, and vice -versa). This 
gives, after some rearrangement and utilizing (\ref{eps_def})
\begin{equation}\label{q_int4}
(1+\epsilon) \ge \frac{1}{g(b/x)} \left( g(a/x) - g(a_{max}/x_{max})\left( 
\frac{x_{max}}{x}\right)^{d-1} \right).
\end{equation}

Inequality (\ref{q_int4}) must be satisfied for all choices of the scaling 
factor $x$. For smaller $x$ ($x \lesssim x_{max}$) $\epsilon$ can be 
negative, but we want to show that as $x$ increases eventually $\epsilon$ 
must become positive. Later we will choose a more restrictive distribution 
of positive energy to better illustrate quantum interest, but first we 
will show that (at least when $d=4$) the total amount of positive energy 
is strictly greater than the total negative energy that passes the 
observer. To do so, evaluate (\ref{q_int4}) in the limit as $x \rightarrow 
\infty$. In this limit for $t=a/x$ and $t=b/x$ we can accurately evaluate 
$g(t)$ in a Taylor series about $t=0$:
\begin{equation}\label{g_taylor}
g(t) = g(0) - \frac{|g''(0)|}{2} t^2 + O(t^4).
\end{equation}
There are no odd powers because of the assumed symmetry in $g$, but even 
if we don't require $g$ to be symmetric there will not be any $t$ term in 
the series because of the peak at $t=0$ (which also forces $g''(0)$ to be 
negative). Thus (\ref{q_int4}) can be written as 
\begin{equation}
\epsilon \ge \frac{|g''(0)|}{2 g(0)} \frac{b^2-a^2}{x^2} - 
\frac{g(a_{max}/x_{max})}{g(0)} \left(\frac{x_{max}}{x}\right)^{d-1} + 
O(1/x^3).
\end{equation} 
In the limit $x\rightarrow\infty$, $\epsilon \rightarrow 0$ and when the 
dimension $d=4$, $\epsilon$ is strictly greater than $0$ for $x$ 
sufficiently large. In 2 dimensional Minkowski space we can only conclude 
that $\epsilon$ is at least zero for arbitrary fluxes using the large $x$ 
behavior of the inequality (\ref{q_int4}). \par
\begin{figure}
\begin{center}
\includegraphics[viewport =36 470 486 710, clip] {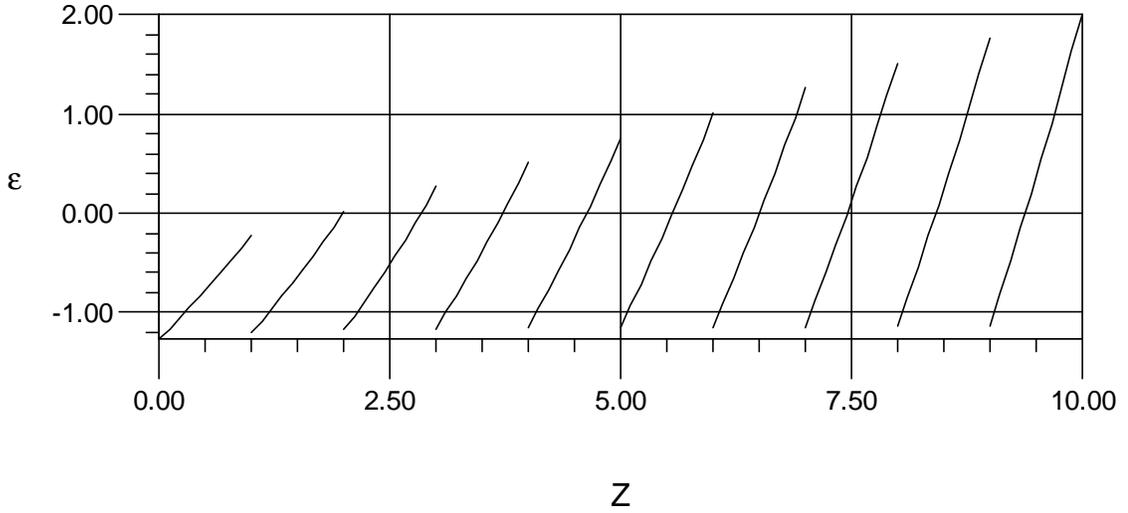}
\end{center}
\caption{\label{e_2d} A lower bound estimate (\ref{q_int_lz}) for the 
quantum interest $\epsilon$ as a function of pulse separation $z=\Delta t 
/ t_0$ in 2D Minkowski spacetime. Ten curves are plotted for values of 
$z_{max}$ from 1 to 10; the range of each curve is $[z_{max}-1,z_{max}]$. 
The larger $z_{max}$ the less negative energy was sampled, allowing 
greater separation of the fluxes. The estimate (\ref{q_int_lz}) is closer 
to the true lower bound for larger $z$. The width of the positive pulse in 
this example is $t_0$, and the sampling function (\ref{power}) with $n=2$ 
was used.} 
\end{figure}
\begin{figure}
\begin{center}
\includegraphics[viewport =36 470 486 710, clip] {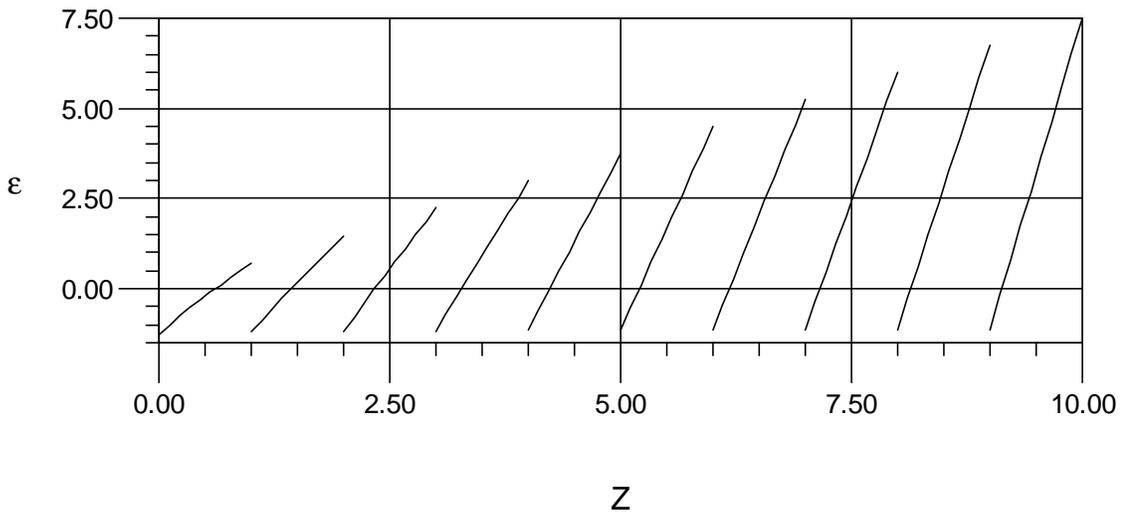}
\end{center}
\caption{\label{e_4d} The same information as shown in Figure (\ref{e_2d}) 
but in 4D Minkowski spacetime.} 
\end{figure}

To gain more insight into inequality (\ref{q_int4}) it is useful to 
restrict the positive flux to last for a time $t_0$. Then 
\begin{equation}\label{r1}
\frac{x t_0}{2}=t_1+t_0
\end{equation}
and
\begin{equation}\label{r2}
\frac{t_0}{2}\le t_1 \le \frac{x_{max} t_0}{2},
\end{equation}
hence
\begin{equation}\label{r3}
3\le x \le x_{max}+2.
\end{equation}
To obtain a lower bound estimate $\epsilon_\ell$ for the quantum interest 
$\epsilon$, set $a=t_0/2$, $a_{max}=0$ and $b=t_1$ in (\ref{q_int4}) (this 
will be a good approximation for larger $x$; see Figure \ref{fig_flux2})
\begin{equation}\label{q_int_l}
1+\epsilon_\ell \ge \frac{g(t_0/2x)-g(0)(x_{max}/x)^{d-1}}{g((t_0/2)(1-
2/x))},
\end{equation}
where we have used (\ref{r1}) to eliminate $t_1$ from the expression. For 
a concrete example we will use the polynomial sampling function with n=2 
(\ref{power}), i.e. $g(t) \propto(t^2-t_0^2/4)^2$. Define $z\equiv 
\frac{\Delta t}{t_0} = \frac{t_1-t_0/2}{t_0}$, so $z$ is the time interval 
separating the positive and negative pulses divided by $t_0$.Using 
(\ref{r1}) to (\ref{r3}) we can find the range of $z$: $0 \le z \le 
z_{max}$, $z_{max}=(x_{max}-1)/2$. When $x=x_{max}$ (and the exact 
inequality (\ref{q_int4}) gives $\epsilon \ge -1$), $z=z_{max}-1$. With 
these definitions (\ref{q_int_l}) becomes (after some simplification)
\begin{equation}\label{q_int_lz}
1+\epsilon_\ell \ge \frac{1}{4}\left[(z+2)^2 - \frac{ (z+3/2)^{5-d} 
(z_{max}+1/2)^{d-1}}{(z+1)^2}\right].
\end{equation}
For large $z$ and $z_{max}$ 
\begin{equation}\label{q_int_lz_ap}
\epsilon_\ell \gtrsim z(\frac{z}{4}+1) -\frac{z^{3-d} z_{max}^{d-1}}{4} 
\left( \frac{3(5-d)-4}{2z}+\frac{d-1}{2 z_{max}} + 1\right).
\end{equation}
When $z$ is in the range $[z_{max}-1,z_{max}]$, (\ref{q_int_lz_ap}) is 
almost a straight line, with $\epsilon_\ell$ ranging from a minimum of $-
5/4$ to a maximum of $z_{max}/4$ in 2D spacetime (compare Figure 
\ref{e_2d} where expression (\ref{q_int_lz}) is plotted), and from $-9/8$ 
to $(3/4) z_{max}$ in 4D spacetime (compare Figure \ref{e_4d}). This shows 
quite clearly that quantum interest grows (almost linearly) as the pulse 
separation increases. But a note of caution: this example will give an 
accurate lower bound on the quantum interest only if our choice of 
sampling function doesn't overestimate the ``real'' $x_{max}$ or $z_{max}$ 
for a given distribution of negative energy. Recall that the ``real'' 
$x_{max}$ must satisfy inequality (\ref{x_max}) for \emph{any} choice of 
sampling function. For example, a sharply peaked sampling function (e.g. 
(\ref{power}) with large $n$) will not give very stringent lower bounds on 
$\rom$, and consequently (\ref{x_max}) will overestimate $x_{max}$ for a 
small pulse of negative energy ($y \ll 1$). A similar analysis to that 
above would then seem to indicate that the quantum interest diverges in 
the limit as $n \rightarrow \infty$ at $z=z_{max}$, but in truth the value 
of $z_{max}$ was overestimated.

\section{\label{sec_massive} Massive scalar fields}
In this section we will briefly show that the quantum interest 
inequalities (\ref{x_max_ie}) and (\ref{q_int4}) and hence all the results 
from the previous section also apply to the massive scalar field in 4 
dimensional Minkowski spacetime. \par
Fewster and Eveson \cite{fe_flat} obtained the following expression for 
$\rom$ in $4D$ Minkowski spacetime for a scalar field of mass $m$:
\begin{equation}\label{fe_rom}
\rom = -A \int_0^{\infty} ds \int_m^{\infty} d \omega_k \ 
\omega_k^2[\omega_k^2-m^2]^{1/2} |\widehat{g^{1/2}}(s+\omega_k)|^2,
\end{equation}
where $A$ is a positive constant, $\widehat{g^{1/2}}(s)$ is the Fourier 
transform of $g^{1/2}(t)$, and one integrates over the spectrum of field 
modes (i.e. $\omega_k=\sqrt{|k|^2+m^2}$, $\vec{k}$ is the 3-momentum of a 
mode with frequency $\omega_k$). \par
If $\rom(m)$ denotes the minimum negative energy bound for a field of mass 
$m$ with sampling function $g(t)$, then
\begin{equation}\label{m_scale}
\robm(m)=\frac{\rom(mx)}{x^4},
\end{equation}
where $\robm(m)$ is the minimum bound with a sampling function 
$\bar{g}(t)=g(t/x)/x$ (the Fourier transform of the scaling relation is 
$\widehat{\bar{g}^{1/2}}(s)=\sqrt{x} \widehat{g^{1/2}}(sx)$). But notice 
from (\ref{fe_rom}) that $\rom(mx) \ge \rom(m)$ for $x\ge 1$ (due to the 
$m$ dependance in the integrand and lower limit of the second integral), 
hence
\begin{equation}\label{m_scale2}
\robm(m) \ge \frac{\rom(m)}{x^4}.
\end{equation}
Thus a massive scalar field will have tighter constraints on allowed 
negative energies than a massless field (compare (\ref{scale})), and all 
the inequalities derived in the previous section remain valid for a 
massive field. (In 2D Minkowski space (\ref{m_scale}) holds with $x^4$ 
replaced by $x^2$, but one cannot conclude that (\ref{m_scale2}) is valid 
$\forall x$.) 

\section{\label{sec_other} Beyond scalar fields in Minkowski spacetime}
The scaling argument used to prove quantum interest for scalar fields 
might readily be applied to other quantum fields, such as the 
Electromagnetic (EM) field or Dirac field, and possibly to certain curved 
spacetimes or Minkowski space with boundary conditions as in the Casimir 
effect. \par
Ford and Roman found a quantum inequality for EM fields in 4D Minkowski 
space using a Lorentzian sampling function \cite{fr_em}:
\begin{equation}\label{qi_em}
\ro_{EM} \ge -\frac{3}{16 \pi^2 t_0^4}.
\end{equation}
This expression certainly indicates that a scaling relation like 
(\ref{scale}) holds for EM fields. The only complication to obtaining 
definitive results in this case is that the Lorentzian sampling function 
does not have compact support, so one cannot rule out the possibility that 
long distance interference effects may spoil quantum interest for 
arbitrary energy fluxes of the EM field (though this seems unlikely). \par
There is some evidence that the Dirac field might also satisfy negative 
energy inequalities similar to those of scalar and EM fields. Vollick has 
recently shown that a superposition of two single particle electron states 
can exhibit negative energy densities, but they are constrained by an 
inequality identical in form to that of the EM and scalar fields 
\cite{vollick}.\par
Fewster and Teo \cite{ft_static} have derived lower bounds of the form 
(\ref{fe_rom}) for states of scalar quantum fields in static, curved 
spacetimes (those with timelike killing vector fields that are 
hypersurface orthogonal). The scaling argument will work in certain static 
spacetimes. For example one can easily show that the scaling relation 
(\ref{m_scale2}) holds in an open static Robertson-Walker universe 
($ds^2=-dt^2+a^2[d\xi^2+sinh^2(\xi)d\Omega^2]$, $a$ is consant), as the 
lower bound for the sampled energy density takes the form 
\cite{ft_static}:
\begin{equation}\label{frw_rom}
\rom = -A \int_0^{\infty} ds \int_C^{\infty} d \omega_k \ 
\omega_k^2[\omega_k^2-C^2]^{1/2} |\widehat{g^{1/2}}(s+\omega_k)|^2,
\end{equation}
where $C=\sqrt{1/a^2+m^2}$ and $m$ is the mass of the scalar field 
(compare (\ref{fe_rom})).\par
In a spacetime with a non-zero expectation value $\rho_0$ for the ground 
state energy density, such as the Boulware state outside a static star or 
with the Casimir effect between two conducting plates, one might expect a 
scaling relation of the form
\begin{equation}
\robm = \frac{\rom}{x^d} + \rho_0
\end{equation} 
to hold. In other words, perhaps one may be able prove the quantum 
interest conjecture for energies \emph{relative} to the ground state 
energy -- see \cite{other_qi} for examples where the quantum inequalities 
take on the from $\ro \ge$ \emph{free field term} + \emph{Casimir 
terms}.\footnote{In fact, such types of inequalities, called `difference 
inequalities', have been derived before in several contexts 
\cite{diff_ie}. I was unaware of these results when I wrote this paper, 
and would like to thank Tom Roman for pointing them out to me.}  

\section{Conclusion}
In this paper we have proven the quantum interest conjecture of Ford and 
Roman for arbitrary distributions of negative energy of scalar fields in 
4D Minkowski spacetime (slightly weaker results hold in 2D). Specifically, 
any flux of negative energy flowing past an inertial observer \emph{must} 
be followed or preceded by positive energy within a finite time interval 
that decreases the larger the amount of negative energy. In addition, the 
total amount of positive energy seen ($E_p$) is always greater than the 
total amount of negative energy ($-|E_m|$). In a more restricted scenario 
where the duration of the positive and negative fluxes are equal, we 
showed that the quantum interest $\epsilon \equiv \left(\frac{E_p}{-|E_m|} 
-1\right)$ grew almost linearly with pulse separation. \par
The nature of existing QI's for EM fields, the Dirac field and scalar 
fields in certain static spacetimes suggests that quantum interest may 
have broader application than free scalar fields in Minkowski spacetime. 
In a situation where the ground state energy density of the field is non-
zero (e.g. in the Casimir effect) we may still expect quantum interest to 
hold, but then ``negative'' energy would refer to energies less than that 
of the ground state.\par
An important consequence of quantum interest is what it tells us about the 
nature of negative energies in free fields. A local pulse of negative 
energy is not an entity that can be manipulated or interacted with 
independently of the accompanying positive energy that must be near by. 
Even if there are states where the positive and negative energies are 
separated by a sizeable distance (as suggested by (\ref{x_max_ie}) when 
the amount of negative energy is very small), one could still only 
interact with the pulse pair as a single entity. For example, absorbing, 
reflecting or scattering only the positive part of the flux would create 
an isolated negative pulse, violating the quantum inequalities. 
Furthermore, this implies that one cannot subject a hot body to a net flux 
of negative energy that otherwise might have lowered its entropy in 
violation of the $2^{nd}$ law of thermodynamics. 
\par

\noindent
{\bf Acknowledgement} \par
I would like to thank Werner Israel for many stimulating discussions. \par

\end{document}